\documentclass[sigconf]{acmart}

\AtBeginDocument{%
  \providecommand\BibTeX{{%
    \normalfont B\kern-0.5em{\scshape i\kern-0.25em b}\kern-0.8em\TeX}}}

\usepackage{balance}
\usepackage{booktabs}
\usepackage{amsmath}
\usepackage{amsfonts}
\usepackage{comment}
\usepackage{footmisc}
\usepackage{mathrsfs}
\usepackage{graphicx}
\usepackage{subfigure}
\usepackage{multirow}
\usepackage{algorithm}
\usepackage{algorithmic}
\usepackage{multicol}
\usepackage{enumitem}
\usepackage{array}
\usepackage{float}
\usepackage{hyperref}
\usepackage[title]{appendix}

\usepackage{xspace}

\newcommand{\ie}{\emph{i.e.,}\xspace}
\newcommand{\name}{DT4IER\xspace}

\setcopyright{acmlicensed}
\copyrightyear{2024}
\acmYear{2024}
\acmDOI{10.1145/3626772.3657829}

\acmConference[SIGIR '24]{Proceedings of the 47th
International ACM SIGIR Conference on Research and Development in
Information Retrieval}{July 14--18, 2024}{Washington, DC, USA}
\acmISBN{979-8-4007-0431-4/24/07}

\begin{document}

\title{Sequential Recommendation for Optimizing Both Immediate Feedback and Long-term Retention}

\author{Ziru Liu}
\affiliation{%
\institution{City University of Hong Kong}
\city{Hong Kong}
\country{China}}
\email{ziruliu2-c@my.cityu.edu.hk}

\author{Shuchang Liu}
\affiliation{%
\institution{Kuaishou Technology}
\city{Beijing}
\country{China}}
\email{liushuchang@kuaishou.com}

\author{Zijian Zhang}
\affiliation{%
\institution{City University of Hong Kong}
\city{Hong Kong}
\country{China}}
\email{zhangzj2114@mails.jlu.edu.cn}

\author{Qingpeng Cai*}
\affiliation{%
\institution{Kuaishou Technology}
\city{Beijing}
\country{China}}
\email{cqpcurry@gmail.com}

\author{Xiangyu Zhao*}
\affiliation{%
\institution{City University of Hong Kong}
\city{Hong Kong}
\country{China}}
\email{xianzhao@cityu.edu.hk}

\author{Kesen Zhao}
\affiliation{%
\institution{City University of Hong Kong}
\city{Hong Kong}
\country{China}}
\email{kesenzhao2-c@my.cityu.edu.hk}

\author{Lantao Hu}
\affiliation{%
\institution{Kuaishou Technology}
\city{Beijing}
\country{China}}
\email{hulantao@kuaishou.com}

\author{Peng Jiang*}
\affiliation{%
\institution{Kuaishou Technology}
\city{Beijing}
\country{China}}
\email{jp2006@139.com}

\author{Kun Gai}
\affiliation{%
\institution{Unaffiliated}
\city{Beijing}
\country{China}}
\email{gai.kun@qq.com}

\thanks{* Corresponding authors.}

\renewcommand{\shortauthors}{Ziru Liu, et al.}

\begin{abstract}
In Recommender System (RS) applications, reinforcement learning (RL) has recently emerged as a powerful tool, primarily due to its proficiency in optimizing long-term rewards. 
Nevertheless, it suffers from instability in the learning process, stemming from the intricate interactions among bootstrapping, off-policy training, and function approximation. Moreover, in multi-reward recommendation scenarios, designing a proper reward setting that reconciles the inner dynamics of various tasks is quite intricate. 
To this end, we propose a novel decision transformer-based recommendation model, DT4IER, to not only elevate the effectiveness of recommendations but also to achieve a harmonious balance between immediate user engagement and long-term retention. The DT4IER applies an innovative multi-reward design that adeptly balances short and long-term rewards with user-specific attributes, which serve to enhance the contextual richness of the reward sequence, ensuring a more informed and personalized recommendation process. To enhance its predictive capabilities, DT4IER incorporates a high-dimensional encoder to identify and leverage the intricate interrelations across diverse tasks. Furthermore, we integrate a contrastive learning approach within the action embedding predictions, significantly boosting the model's overall performance. Experiments on three real-world datasets demonstrate the effectiveness of DT4IER against state-of-the-art baselines in terms of both immediate user engagement and long-term retention. The source code is accessible online to facilitate replication \footnote{https://github.com/Applied-Machine-Learning-Lab/DT4IER}.

\end{abstract}



\maketitle

\section{Introduction}

In today's digital age, platforms spanning from social media to e-commerce have led to an explosion in the amount of information available online, underscoring the importance of efficient navigation tools~\cite{aceto2020industry}. 
Recommender Systems (RS) have emerged as a crucial technology in this realm, adeptly improving content suggestions and optimizing recommendations according to user interests inferred from their historical engagement.
In recent years, researchers have proposed various methods for Recommender Systems, encompassing collaborative filtering ~\cite{mooney2000content}, matrix factorization-based approaches ~\cite{koren2009matrix}, and those powered by deep learning ~\cite{cheng2016wide,zhang2019deep, chen2022knowledge}. 
Among them, transformer-based models, notably BERT4Rec~\cite{sun2019bert4rec} and SASRec~\cite{kang2018self} have risen to prominence, redefining the landscape of recommendation systems. 
It is believed that the strength of transformers lies in the attention mechanisms~\cite{46201}, which can dynamically capture the inherent dependencies in data, offering an accurate understanding of user patterns. 
This adaptability and precision make transformers well-suited for navigating the realm of user preferences inference and immediate feedback prediction.

In practice, a range of metrics such as clicks, likes, and ratings are widely used as indicators of user preferences.
Though effective, these immediate metrics offer limited insights into the user's lasting impression in the long term.
In some cases, they can be misleading indicators of content quality~\cite{wu2017returning,yi2014beyond}. 
For instance, content with an eye-catching title but poor content quality may initially draw attention and obtain positive feedback, but only to later abuse the users' trust and undermine their loyalty. 
This highlights a pressing need to balance the RS's focus between short-term user engagement and long-term user satisfaction~\cite{viljanen2016modelling}
.
To this end, methods based on Reinforcement Learning (RL) have emerged~\cite{chen2021survey}.
By modeling sequential user behaviors using a Markov Decision Process (MDP), RL-based systems can dynamically adapt recommendations at every user interaction juncture~\cite{zhao2011reinforcement, mahmood2007learning}. 
The strength of RL lies in its capability to optimize cumulative rewards over extended sequences in the future, allowing it to capture long-term objectives and trends, and ensure sustained user engagement.
A recent finding~\cite{cai2023reinforcing} shows that RL-based RS can even directly optimize user retention, a crucial metric often ignored but paramount in real-world business contexts~\cite{zou2019reinforcement, chen2023sim2rec} that is closely related to daily active users (DAU).

Despite their promises, RL-based systems have some inherent challenges. 
Firstly, the long-term credit assignment through bootstrapping leads to the \textit{Deadly Triad} issue which emerges from the complex interplay between bootstrapping, off-policy training, and function approximation, often making the learning process unstable~\cite{chen2019generative}. 
Secondly, the standard practice of discounting future rewards in Temporal Difference (TD) learning can inadvertently drive the system to be overly focused on immediate gains at the expense of long-term objectives \cite{xu2018meta}. 
To circumvent these problems and unlock the full potential of RL-based recommendation systems, the innovative Decision Transformer (DT)~\cite{chen2021decision} has been introduced and then applied in RS. 
During inference, the recommendation policy is conditioned on the rewards, i.e. returns-to-go (RTG).
During training, it reformulates the reinforcement learning paradigms into sequence modeling tasks, and the transformer model is used to capture not only the interactions but also the user rewards.
With this transformation, DT converts the intricate landscape of RL into a tractable one close to supervised learning.
Additionally, DT has also been proven to be efficient in boosting recommendation performance with respect to user retention~\cite{zhao2023user}.

We posit that focusing solely on immediate user feedback or long-term retention is insufficient. A more holistic approach requires optimizing both metrics simultaneously, offering a comprehensive perspective on user behavior. This approach frames the problem as a long short-term multi-task learning challenge.
Adapting Decision Transformers (DT) to multi-reward contexts, however, presents significant challenges.
Firstly, the intricate dynamics among various user responses in historical data underscore the complexity involved in designing corresponding rewards. 
This is crucial as it directly impacts the quality of the recommendations~\cite{ratner2018simplifying}. 
Secondly, the presence of multiple objectives introduces additional complexity to the training process. 
This arises from the sophisticated and often unpredictable interdependencies among different tasks, potentially impairing the model's learning efficiency.

To address these challenges, we introduce \textbf{\name}, a novel framework based on the Decision Transformer, crafted for long short-term multi-task recommendation scenarios. 
Our approach deploys an innovative multi-reward configuration, reinforced by a high-dimensional encoder designed to capture the intricate relationships among different tasks effectively. 
Furthermore, we apply an innovative approach to reward structuring, skillfully balancing short-term and long-term rewards. It does so by incorporating user-specific attributes, which serve to enhance the contextual richness of the reward sequence ensuring a more informed and personalized recommendation process.
We also introduce a contrastive learning objective to ensure that the predicted action embeddings for distinct rewards do not converge too closely.
Our key contributions in this paper can be summarized as follows:
\begin{itemize}[leftmargin=*]
    \item We emphasize the importance of long short-term multi-task sequential recommendations and introduce \name, a novel Decision Transformer-based model engineered for integrated user engagement and retention.
    
    \item Our innovative framework applies a novel multi-reward setting that balances immediate feedback with long-term retention by user-specific features, and then complements by a corresponding high-dimensional embedding module and a contrastive loss term.
    \item We validate the performance of \name through extensive experimentation compared with state-of-the-art Sequential Recommender Systems (SRSs) and Multi-Task Learning (MTL) models on three real-world datasets.
\end{itemize}


\section{Preliminaries}
In this section, we provide an overview of the foundational concepts and primary notations used throughout this paper. 
\subsection{Offline Reinforcement Learning}
Given a Markov decision process (MDP) formulated as $(\mathcal{S}, \mathcal{A}, P, \mathcal{R}, \gamma)$ where $\mathcal{S}$ is the set of state $s \in \mathbf{R}^d$, $\mathcal{A}$ is the set of action $a$, $P(s'|s,a)$ is the transition probabilities from state $s$ to new state $s'$ given action $a$, $\mathcal{R}$ is the reward function for specific state-action pairs and $\gamma$ is the discount rate. The trajectory from timestamp $0$ to $T$ can be written as $(s_0,a_0,r_0,\cdots,s_t,a_t,r_t,\cdots,s_T,a_T,r_T)$, where $(s_t,a_t,r_t)$ are state, action and reward pairs at timestamp $t$. The learning objective of RL is to determine an optimal policy that maximizes the expected cumulative return $\mathbb{E}\left[\sum_{t=1}^T \gamma^t r_t\right]$ given a specific reward function and discount rate.

\subsection{Decision Transformer}
\label{section:DT}
Rather than employing traditional RL algorithms such as training an agent policy or approximating value functions \cite{48200}, the decision transformer follows a different routine that recasts RL into a sequence modeling problem with supervised learning objectives. To equip the transformer with the capability to discern significant patterns, the corresponding trajectory representation with $T$ timestamps is designed as:
\begin{equation}
    \tau=\left(\widehat{R}_1, s_1, a_1, \ldots, \widehat{R}_t, s_t, a_t, \ldots, \widehat{R}_T, s_T \right)
\end{equation}

In this representation, 
$\widehat{R}_t = \sum_{k=t}^T r_k$ defines the returns-to-go (RTG) which represents the cumulative reward from time $t$ through to time $T$, without applying any discount. Given the RTG and state information, the DT is capable of predicting the next action $a_T$ by a causally masked transformer architecture with layered self-attention and residual connections \cite{chen2021decision}.
In each layer, $m$ input embeddings, denoted as \{$x_i$\}$_{i=1}^{m}$, are processed to yield corresponding output embeddings \{$z_i$\}$_{i=1}^{m}$. Each token's position, $i$, dictates its transformation into a key $k_i$, a query $q_i$, and a value $v_i$ \cite{46201}. The resultant output for the same position is derived by adjusting values, $v_j$, using the normalized dot product between the query $q_i$ and its associated keys $k_j$ which is further processed by a softmax activation function $\sigma_s$:
\begin{equation}
    z_i=\sum_{j=1}^m \sigma_s\left(\left\{\left\langle q_i, k_l\right\rangle\right\}_{l=1}^m\right)_j \cdot v_j
\end{equation}
In the realm of recommendation systems, the inference process utilizing the DT can be concisely described as follows: Initially, the model receives a combination of state and action inputs. These inputs are processed through the Decision Transformer, resulting in an output that consists of an action embedding specifically prompted by a designed reward which is integral to guiding the model's decision-making process. Subsequently, this action embedding undergoes a decoding process, which ultimately yields a sequence of recommended items.

\begin{figure*}
    \centering
    \includegraphics[width=0.95\linewidth]{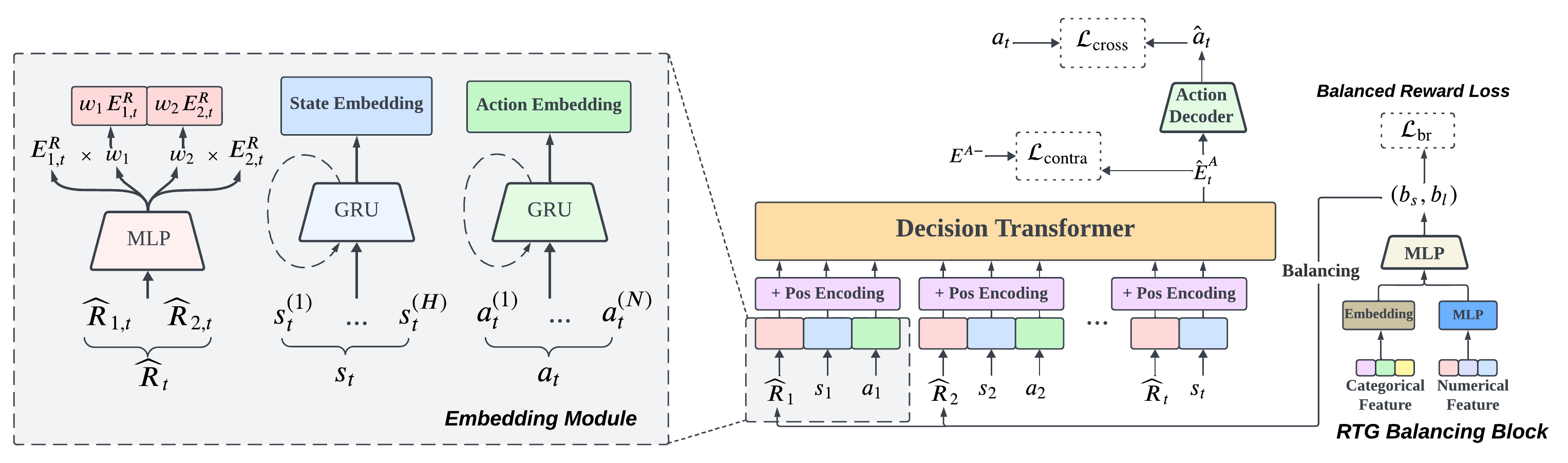}
    \vspace{-5mm}
    \caption{Overview Framework of \name.}
    \label{fig:Frame}
    \vspace{-4mm}
\end{figure*}

\section{THE PROPOSED Framework}
In the following section, we provide a comprehensive overview of our proposed method, \name. We begin with the problem formulation, setting the stage for a deeper understanding, and then detailing the intricate design and settings of the specific modules.

\subsection{Problem Defination}
In the conventional sequential recommendation scenario, the objective is to recommend items based on a user's historical sequence optimized for specific indicators. However, from a business perspective, focusing on a single metric can be limiting since user behavior can be influenced by various factors and can exhibit different patterns over time. Therefore, we prioritize the optimization of two key performance indicators: click-through rate (CTR) and return frequency. The former serves as a widely recognized business metric in numerous real-world applications, offering immediate insight into user engagement. The latter is intrinsically tied to vital operational metrics, including daily active users (DAU), which are essential for sustained platform growth and user retention. This optimization strategy is applicable to a variety of digital platforms, encompassing streaming services, e-commerce websites, and social media networks, where both immediate user engagement and long-term user retention are critical for success. To realize this, we adapt the DT framework to a multi-reward setting, whose architecture can be naturally extended to handle multiple reward signals. Given the input trajectory $\tau=\left(\widehat{\textbf{R}}_1, \textbf{s}_1, \textbf{a}_1, \ldots, \widehat{\textbf{R}}_t, \textbf{s}_t, \textbf{a}_t, \ldots, \widehat{\textbf{R}}_T, \textbf{s}_T \right)$, the state-action and RTG is defined as:
\begin{itemize} [leftmargin=*]
    \item \textbf{Session} $t$ represents an individual timestamp within a trajectory of length $T$. It also signifies a specific day for a particular user.
    \item \textbf{State} $\textbf{s}_t \in \mathbf{R}^H$ represents the historical interaction information for a user before session $t$, which comprises user-clicked item IDs, zero-padded to length $H$. It will be updated based on clicked items in the current action.
    \item \textbf{Action} $\textbf{a}_t \in \mathbf{R}^N$ is the recommendation list containing item IDs of length $N$, denoting the action based on state $s_t$. 
    \item \textbf{Reward} $\textbf{r}_t = (r_{s,t},r_{l,t}) \in \mathbf{R}^2$ represents the feedback corresponding to the actions $a_t$ executed at session $t$. In this paper, our focus is primarily on two pivotal metrics in real-world recommender systems. The short-term indicator $r_{s,t}$ is quantified as the \textbf{click-through rate} for recommended action $a_t$ at session $t$. And long-term indicator $r_{l,t}$ is defined as the \textbf{return frequency} for a given session $t$ to measure user retention.
    \item \textbf{Return-to-go (RTG)} $\widehat{\textbf{R}}_t \in \mathbf{R}^2$ denote the accumulative reward accumulated from session $t$ through to $T$, without the introduction of any discount factor, which can be expressed as:
    \begin{equation}
        \widehat{\textbf{R}}_t = [\widehat{R}_{s,t},\widehat{R}_{l,t}] = \left[ \sum_{i=t}^T r_{s,i} \ ,\ \sum_{i=t}^T r_{l,i} \right]
    \end{equation}
\end{itemize}

Our method aims to optimize a spectrum of objectives that encompass both short-term and long-term rewards. This approach can be regarded as both an evolution and a specialization within the MTL framework, one that is finely tuned to balance and achieve both immediate and enduring user engagement metrics.

\subsection{Framework Overview}
The overall structure of \name is illustrated in Figure \ref{fig:Frame}. While it shares foundational elements with the conventional DT~\cite{chen2021decision}, our design incorporates modifications designed to accommodate recommendation tasks with a multi-task setting. Here's a step-by-step overview of \name's forward process:
\begin{itemize}[leftmargin=*]
    \item \textbf{Adaptive RTG Balancing Block}: To effectively capture the intricate dynamics of short-term and long-term user behaviors, we propose to apply user feature-based reweighting to the RTG sequence. Further insights can be found in Section \ref{section:RR}.
    \item \textbf{Embedding Module}: The modified trajectory data then undergoes an embedding transformation to be represented as dense vectors. To be specific, the state and action sequences are processed using a Gated Recurrent Unit (GRU) \cite{cho2014learning}, detailed further in Appendix \ref{S-A E}. In contrast, the reward sequence employs a high-dimensional encoder which is further explained in Section \ref{section:RE}. Additionally, a positional session encoding is integrated.
    \item \textbf{Transformer Decision Block}: Upon processing, the dense trajectory vectors serve as context, guiding the generation of action embeddings for the subsequent timestamp. This decision-making module is underpinned by the transformer model, detailed further in the Appendix \ref{TB}.
    \item \textbf{Action Decoding Block}: Using the predicted action embeddings, our decoding unit strives to construct an action sequence that aligns closely with the reference of ground truth actions, detailed further in Appendix \ref{AD}.
    \item \textbf{Training Objective}: We employ an objective function designed to minimize the differences between the predicted and ground truth actions. Besides, a supplementary contrastive learning objective is introduced, serving as a catalyst to amplify the model's robustness. More details are given in Section \ref{section:CL}.
\end{itemize}

\subsection{Adaptive RTG Balancing}
\label{section:RR}
In settings that involve long short-term reward optimization, striking a balance between short-term and long-term performance metrics is notably challenging. This complexity arises from inherent conflicts between immediate and deferred objectives, as well as from the convoluted interdependencies among tasks \cite{ratner2018simplifying}. To adeptly balance this equilibrium, we propose a novel strategy: the balancing of the RTG sequence, adaptively controlled by distinct user features. Since these user features hold a wealth of information about user preferences, they serve as a reliable guide for understanding user behaviors, bridging the gap between their immediate feedback and long-term retention.

Considering the diverse nature of user features, it is crucial to treat numerical and categorical data distinctly. We process categorical features through an embedding layer to capture their unique attributes effectively. Meanwhile, numerical features are refined using a Multilayer Perceptron (MLP). The outputs of both processes are then combined and fed into another MLP which is designed to refine the data further and produce tailored weights that strike a balance between immediate responses and long-term engagement in user behavior. Specifically, given a set of user features $\textbf{U} = (\mathbf{u}_\text{n},\mathbf{u}_\text{c})$, where $\mathbf{u}_\text{n}$ are numerical features and $\mathbf{u}_\text{c}$ are categorical features. The whole process can be summarized as:

\begin{itemize} [leftmargin=*]
    \item \textbf{Feature Extraction:} The transformation of categorical features into an embedded space is given by:

\begin{equation}
    \mathbf{e}_i = E(c_i), \quad \forall c_i \in \mathbf{u}_\text{c}
\end{equation}

where $E$ represents the embedding function, and $\mathbf{e}_i$ is the embedding for the $i$-th categorical feature. The embedded vectors are then concatenated to form $\mathbf{u}_\text{e} = [\mathbf{e}_1, \mathbf{e}_2, \ldots, \mathbf{e}_n]$.

\noindent The numerical features are transformed by an MLP with the sigmoid activation function to obtain:

\begin{equation}
    \mathbf{u}_\text{M} = \text{MLP}_1(\mathbf{u}_\text{num})
\end{equation}

\noindent To derive the long short-term balancing weights, the combined feature vector $\mathbf{u}_\text{com} = [\mathbf{u}_\text{e}, \mathbf{u}_\text{M}]$ is processed by another MLP with softmax activation function to yield a set of weights $\mathbf{B} = (b_s,b_l)$ which indicates the importance of each reward indicator:

\begin{equation}
    \mathbf{z} = \text{MLP}_2(\mathbf{u}_\text{com})
\end{equation}
\begin{equation}
    \mathbf{B} = \text{softmax}(\mathbf{z}) = \left[ \frac{e^{z_i}}{\sum_{j=1}^{K} e^{z_j}} \right]_{i=1}^{K=2}
\end{equation}

where $z_i$ represents the $i$-th element of the output vector $\mathbf{z}$, and $K$ is the total number of elements in $\mathbf{z}$, representing weights for balancing immediate response and long-term retention.
    \item \textbf{Balanced Reward Loss:} To guide the user-specific weight optimization process, we hope the learned balancing weights can maximize the overall weighted sum of immediate response reward $r_{s,i}$ and long-term retention reward $r_{l,i}$ and also balance them to reach a better balance. This objective can be written as:
    \begin{eqnarray} \label{eq:O}
        \mathcal{O} 
        &=& \sum_{k=1}^K \left[ \sum_{i=1}^{I_k} [(b_s \cdot r_{s,i} + b_l \cdot r_{l,i}) - \left\|b_s \cdot r_{s,i}-b_l \cdot r_{l,i}\right\|_2^2 ] \right]  \nonumber \\
        &\leq& \sum_{k=1}^K [(b_s \cdot \widehat{R}_{s,k} + b_l \cdot \widehat{R}_{l,k}) - \left\|b_s \cdot \widehat{R}_{s,k}-b_l \cdot \widehat{R}_{l,k}\right\|_2^2] \\
        \nonumber
    \end{eqnarray}
    where $K$ is the number of users, $I_k$ is the number of interacted items for user $k$. $b_s$ and $b_l$ represent the weights for immediate feedback and long-term retention with condition $b_s + b_l = 1$.

    To achieve a balanced consideration of rewards for immediate feedback and long-term retention, we implement an $L_2$ regularization approach which imposes a penalty on substantial deviations between the weighted values of short-term and long-term rewards. Such a strategy is designed to prevent the system from disproportionately favoring one type of reward over the other, thus maintaining an equitable focus on both immediate and sustained user engagement. Then the second inequality in Equation (\ref{eq:O}) is derived from the fact that RTG is the sum of rewards. 
    
    To maximize the objective $\mathcal{O} $, we design the corresponding \textit{BalancedRewardLoss} function as follows:
    \begin{equation}
        \mathcal{L}_{br} = -\sum_{k=1}^K [(b_s \cdot \widehat{R}_{s,k} + b_l \cdot \widehat{R}_{l,k}) - \gamma \left\|b_s \cdot \widehat{R}_{s,k}-b_l \cdot \widehat{R}_{l,k}\right\|_2^2]
    \end{equation}
    where $\gamma$ is a hyperparameter to control the balance term.
    \end{itemize}

The overall design of the \textit{BalancedRewardLoss} function aligns with the objective of achieving a sustainable and effective recommendation strategy, addressing both immediate user responses and long-term user retention. The existing RTG sequence is further rebalanced by the weights $B$ and we also use the same notation for the RTG sequence $\widehat{\textbf{R}}_t$.

\subsection{Multi-reward Embedding}
\label{section:RE}
In the architecture of DT, the reward mechanism is crucial as it drives the process of predicting actions. The reward signals to the model what outcomes to strive for, influencing its predictive decisions. However, the use of a simplistic embedding strategy for representing these rewards falls short, as it fails to preserve the essential partial order relationships among them. This limitation is particularly acute in scenarios involving multiple rewards, where the model must understand and respect the hierarchy of rewards to make accurate and contextually relevant embeddings. In order to effectively solve this problem, we introduce an innovative multi-reward embedding module specifically tailored for our RTG setting. This method employs learnable weights derived from an MLP, anchoring the weights directly to the reward values. Such an approach not only allows for a more nuanced representation of rewards but also ensures that the model remains adaptive to shifts in user behavior patterns. This process can be delineated as follows:
\begin{itemize} [leftmargin=*]
    \item \textbf{Discretization of Rewards} With the given numerical reward value $\widehat{\textbf{R}}_t = [\widehat{R}_{s,t},\widehat{R}_{l,t}]$ as a starting point, we apply a discretization technique enabling us to extract distinct meta-embeddings $\textbf{E}^R = [E^R_{s,t}, E^R_{l,t}]$ tailored to each task's reward. The principle behind this is to transform continuous reward values into categorical bins, each associated with a specific embedding, allowing for more subtle representations.
    \item \textbf{Weighted Score Generation} The reward values, once processed, are channeled into an MLP with a specific architecture, which translates the reward value into a multi-weighted score. To be specific, the MLP layer can be formulated as follows:
    \begin{equation} \label{eq:mlp} 
    \begin{aligned}
        \textbf{h}_{n+1} = \sigma(\textbf{W}_n\textbf{h}_n + \textbf{b}_n) \quad ,n=0,1,...,N-2  \\
        \textbf{h}_{N} = \sigma^*(\textbf{W}_{N-1}\textbf{h}_{N-1} + \textbf{b}_{N-1}) \qquad \qquad
    \end{aligned}
    \end{equation}
    where $\textbf{h}_n$ represent the $n$-th hidden layer, characterized by its weight $\textbf{W}_n$ and bias $\textbf{b}_n$. For this layer, the activation function used is Leaky-ReLU, denoted as $\sigma$. Meanwhile, the output layer, symbolized by $\textbf{h}_N$, employs the softmax function $\sigma^*$. Here the output of MLP is a 2-D vector with $\textbf{w} = \textbf{h}_{N} = [w_1,w_2]$.
    The core purpose of this step is to harness the potential of deep learning to derive a relational significance score for each task, highlighting the underlying relationship between rewards.
    
    \item \textbf{Embedding Concatenation} Upon obtaining the weighted meta-embeddings specific to each task, we proceed to concatenate them to ensure a unified, comprehensive reward representation, providing a holistic view of the reward dynamics across tasks:
    \begin{equation}
        \widehat{\textbf{E}}^R = concate(w_1E^R_{s,t}, w_2E^R_{I,t})
    \end{equation}
    where $concate()$ represents the concatenation operation.
\end{itemize}

\noindent This embedding module corresponds to the shared representation in MTL, drawing from both types of rewards to offer a comprehensive understanding of the tasks. Thus, while not being classic MTL, our approach borrows the foundational principle of jointly optimizing for multiple objectives to enhance overall performance.

\subsection{Objective Function with Contrastive Learning Term}
\label{section:CL}
While DT typically predicts actions based on the prospect of obtaining the maximum reward, this may inadvertently sideline data samples associated with lower rewards \cite{zhao2023user}. Furthermore, for optimal performance, it's essential that actions with different reward values are distinctly separable in the embedding space. To achieve this, we introduce a contrastive learning-based loss term, ensuring the model effectively differentiates between actions corresponding to varied rewards. 
It can be written as:
\begin{equation}
    \mathcal{L}_{contra} = -\sum_{\widehat{\textbf{E}}^{A-} \in \Omega^{-}} D(\widehat{\textbf{E}}^A,\widehat{\textbf{E}}^{A-})
\end{equation}

\noindent where $\widehat{\textbf{E}}^{A}$ is the predicted action embedding, $\Omega^{-}$ is the set for negative samples $\widehat{\textbf{E}}^{A-}$, and function $D(x,y)$ calculates the similarity of two sequences. In our context, negative samples are identified as data points where the rewards $r_{1,t}$ and $r_{2,t}$ are consistently below 0.6. This threshold signifies that these samples underperform, falling below average in both click rate and retention metrics. 

Besides, in the original Decision Transformer model, the $L_2$ loss is employed for scenarios involving a continuous action space. However, in our specific setting, the action comprises video IDs with a length of 30, representing a discrete space. To adapt to this context, each element within the action space is converted into a one-hot encoded label. Consequently, we utilize the cross-entropy loss function, which effectively measures the divergence between the predicted action distribution and the ground truth action:
\begin{equation}
    \mathcal{L}_{cross} = -\sum_{z=1}^{Z} y_{o, z} \log \left(p_{o, z}\right)
\end{equation}

\noindent where $Z$ is the length of the action sequence, $y_{o,z}$ is the binary indicator equal to 1 if the current predicted action $o$ is inside the action table with label $z$, $p_{o,z}$ is predicted probability for current predicted action $o$ is of class $z$. 
Ultimately, the overall objective function can be expressed as:
\begin{equation}
    \mathcal{L} = \mathcal{L}_{cross} + \alpha \mathcal{L}_{contra}
\end{equation}
where $\alpha$ is the contrastive learning loss weight. 




\section{Experiment}
In this section, we assess the performance of the \name framework using experiments conducted on two real-world datasets.

\subsection{Dataset}
We carried out our experiment using three datasets.
\begin{itemize} [leftmargin=*]
    \item \textbf{Kuairand-Pure} \footnote{https://kuairand.com/} is an unbiased sequential recommendation dataset featuring random video exposures. 

    \item \textbf{MovieLens-25M} \footnote{https://grouplens.org/datasets/movielens/25m/}, a widely-used benchmark for SRSs, boasts a more extensive scale but with a sparser distribution. 

    \item \textbf{RetailRocket}\footnote{https://www.kaggle.com/datasets/retailrocket/ecommerce-dataset} dataset is collected from a real-world e-commerce website. To optimize the transformer's memory requirements, item IDs have been reindexed.
\end{itemize}

\begin{table*}
\centering
\caption{Overall Performance on three datasets for different models.}
\vspace{-3mm}
\begin{tabular}{@{}|l|l|ccccccc|@{}}
\toprule
\multirow{2}{*}{Dataset} & \multirow{2}{*}{Metric} & \multicolumn{7}{c|}{Model} \\
\cmidrule(l){3-9}
 & & {MMoE} & {PLE} & {RMTL-PLE} & {BERT4Rec} & {SASRec} & {DT4Rec} & {\name} \\
\midrule
\multirow{5}{*}{Kuairend-Pure} & BLEU & 0.654 & 0.658 & 0.674 & 0.781 & 0.787 & \underline{0.821} & \textbf{0.892*} \\
 & ROUGE & 0.889 & 0.888 & 0.891 & 0.855 & 0.874 & \underline{\textbf{0.911}} & 0.89 \\
 & HR & 0.655 & 0.657 & 0.668 & 0.681 & 0.687 & \underline{0.731} & \textbf{0.838*} \\
 & NDCG & 0.531 & 0.534 & 0.552 & 0.690 & 0.732 & \underline{0.754} & \textbf{0.852*} \\
 & SB-URS & 411,278 & 412,359 & 432,402 & 507,831 & 512,490 & \underline{526,032} & \textbf{529,741} \\
\midrule
\multirow{5}{*}{ML-25M} & BLEU & 0.268 & 0.27 & 0.273 & 0.476 & 0.479 & \underline{0.560} & \textbf{0.594*} \\
 & ROUGE & 0.264 & 0.268 & 0.269 & 0.324 & 0.331 & \underline{0.388} & \textbf{0.415*} \\
 & HR & 0.267 & 0.269 & 0.272 & 0.281 & 0.287 & \underline{0.334} & \textbf{0.401*} \\
 & NDCG & 0.281 & 0.284 & 0.286 & 0.322 & 0.338 & \underline{0.361} & \textbf{0.418*} \\
 & SB-URS & 114,231 & 115,087 & 115,842 & 135,928 & 141,823 & \underline{151,302} & \textbf{153,135} \\
\midrule
\multirow{5}{*}{RetailRocket} & BLEU & 0.268 & 0.273 & 0.294 & 0.455 & 0.457 & \underline{0.572} & \textbf{0.628*} \\
 & ROUGE & 0.872 & 0.870 & 0.878 & 0.892 & 0.898 & \underline{0.889} & \textbf{0.908*} \\
 & HR & 0.371 & 0.377 & 0.384 & 0.393 & 0.398 & \underline{0.411} & \textbf{0.439*} \\
 & NDCG & 0.411 & 0.415 & 0.431 & 0.45 & 0.461 & \underline{0.487} & \textbf{0.544*} \\
 & SB-URS & 187,943 & 188,462 & 193,210 & 201,764 & 204,218 & \underline{225,628} & \textbf{238,338} \\
\bottomrule
\end{tabular}
\label{table:2}
\vspace{2mm}
\\``\textbf{{\Large *}}'': the statistically significant improvements (\ie two-sided t-test with $p<0.05$) over the best baseline.
\\ \underline{Underline}: the best baseline model. \textbf{Bold}: the best performance among all models.
\vspace{-2mm}
\end{table*}

\subsection{Evaluation Metrics}
We evaluate \name's effectiveness using various metrics, focusing on both short-term recommendation accuracy and long-term user retention. Detailed metrics are provided below:
\begin{itemize} [leftmargin=*]
    \item \textbf{BLEU} \cite{10.3115/1073083.1073135} assesses the precision of the predicted recommendation list which is a common metric in SRSs.
    \item \textbf{ROUGE} \cite{lin-2004-rouge} calculates the recall rate of the predicted recommendation list.
    \item \textbf{HR@K} quantifies the likelihood of ground-truth items ranking within the top-K recommendations.
    \item \textbf{NDCG@K} \cite{wang2013theoretical} calculates the normalized cumulative gain within the top-K recommendations, factoring in positional relevance.
    \item \textbf{Similarity-Based User Return Score} (SB-URS) \cite{zhao2023user} is an established metric for assessing the retention impact of recommended lists, determined by the weighted sum of the actual user retention score. In our approach, we categorize samples into eight distinct classes based on their reward values, uniformly ranging from 0 to 1. The similarity, represented by the BLEU score, is then computed by comparing the predicted recommendations with the ground truth for each class:
    \begin{equation}
        \mathrm{SB}-\mathrm{URS}=\sum_{c=0}^7 sim_c \cdot\left(r_c-\frac{1}{2}\right) \cdot N_c
    \end{equation}
    Here, $sim_c$ represents the similarity for class $c$, $r_c$ denotes the corresponding ground truth retention reward, and $N_c$ signifies the count of samples with a reward classification of $c$.
\end{itemize}

\subsection{Baselines}
In our analysis, we contrast the performance of our approach with state-of-the-art (SOTA) models spanning both multi-task learning and decision transformer-based sequential recommendations:

\begin{itemize} [leftmargin=*]
    \item \textbf{MMoE} \cite{ma2018mmoe}: Renowned as a robust multi-task learning model, MMoE employs gating mechanisms to manage the interplay between the shared foundation and task-specific layers.
    \item  \textbf{PLE} \cite{tang2020ple}: Standing out as a leading multi-task learning solution, PLE handles complex task interrelations by employing a blend of shared experts and task-specific expert layers.
    \item \textbf{RMTL} \cite{liu2023multi}: This is a reinforcement learning-based multi-task recommendation model with adaptive loss weights.
    \item \textbf{BERT4Rec} \cite{sun2019bert4rec}: It employs a bidirectional Transformer architecture to effectively capture sequential patterns in user behavior with a masked language model.
    \item \textbf{SASRec} \cite{kang2018self}: This model applies a left-to-right unidirectional Transformer to capture user preference.
    \item \textbf{DT4Rec} \cite{zhao2023user}: Operating on the decision transformer paradigm, this Sequential Recommendation System (SRS) model is meticulously tailored for optimizing user retention.
\end{itemize}

\noindent Furthermore, we've adapted the model architectures of MMoE and PLE to facilitate sequential recommendations, utilizing a weighted score derived from multiple tasks. 

\subsection{Implementation Details}
For the Decision Transformer model, we use a trajectory length $T$ of 20 for both datasets, and we set the maximum sequence length the same for state and action $H=N=30$. The model configuration includes 2 Transformer layers, 8 heads, and an embedding size of 128. We employ the Adam optimizer, with a batch size set to 128. The learning rates are 0.005 for Kuairand-Pure and 0.02 for ML-25M and RetailRocket. The action decoder is capped at a maximum sequence length of 30. The balanced reward loss utilizes a balance term $\gamma$ of 0.5, and a contrastive loss parameter $\alpha$ of 0.1. For other baseline models, we either adopt the optimal hyper-parameters suggested by their original authors or search within the same ranges as our model. All results are showcased using the optimal configurations for each model, as detailed in Appendix \ref{AppendixA}.


\subsection{Overall Performance and Comparison}
We assessed the efficacy of our proposed \name model against four baselines across two datasets. A comprehensive performance summary is shown in Table \ref{table:2}, yielding the following insights:

\begin{itemize} [leftmargin=*]
    \item \textbf{Limitations of MMoE}: Among all models, MMoE demonstrates the least satisfactory performance concerning both recommendation accuracy and long-term retention across the datasets. While MMoE's design proficiently handles multiple tasks by balancing parameter interactions between shared components and task-specific towers, its architecture, optimized for parallel task processing, struggles to adapt to the dynamic and evolving nature of sequential data. Similarly, PLE faces the same challenge, potentially resulting in diminished outcomes in sequential recommendation contexts.
    \item \textbf{Strengths of DT4Rec}: DT4Rec achieves the best performance in both recommendation accuracy and retention among all baseline models. Its unique auto-discretized reward prompt design guides the model training towards boosting long-term user engagement, thus enhancing user retention appreciably.
    \item \textbf{Superiority of \name}: The \name model consistently outperforms the four baselines in both the realms of recommendation accuracy and user retention score across datasets. Particularly on the Kuairand-Pure dataset, \name exhibits 0.07-0.09 improvement in recommendation accuracy compared to the top-performing baseline. By interpreting RL as an autoregressive setting and integrating a structure designed for multiple rewards, our model strikes a balance between immediate feedback and long-term retention, resulting in significant enhancements in recommendation performance while retaining user engagement.
    
\end{itemize}
To conclude, the \name model represents a significant advancement over existing state-of-the-art multi-task learning (MTL) frameworks and transformer-based sequential recommendation systems. It excels in providing superior recommendation accuracy and sustaining long-term user retention across diverse real-world datasets. This is largely due to its sophisticated reward structure, which has been meticulously designed to harmonize the delivery of immediate feedback and long-term retention.

\subsection{Ablation Study}

In this subsection, we delve into an ablation study to underscore the significance of the distinct modules integrated within our proposed model. By contrasting variants of the primary model with specific modules omitted, we aim to measure the impact of each component. The variant models are delineated as follows:

\begin{itemize} [leftmargin=*]
    \item \textbf{NAW} represents the model variant devoid of the adaptive RTG weighting module, with all other components kept constant.
    \item \textbf{NRE} In this configuration, a standard reward embedding is employed without considering intricate task relations.
    \item \textbf{NCL} This variant exclusively leverages the cross-entropy loss in its objective functions without the contrastive loss component.
\end{itemize}

\begin{figure}[h] 
    \centering
    \vspace{-3mm}{\subfigure{\includegraphics[width=0.325\linewidth]{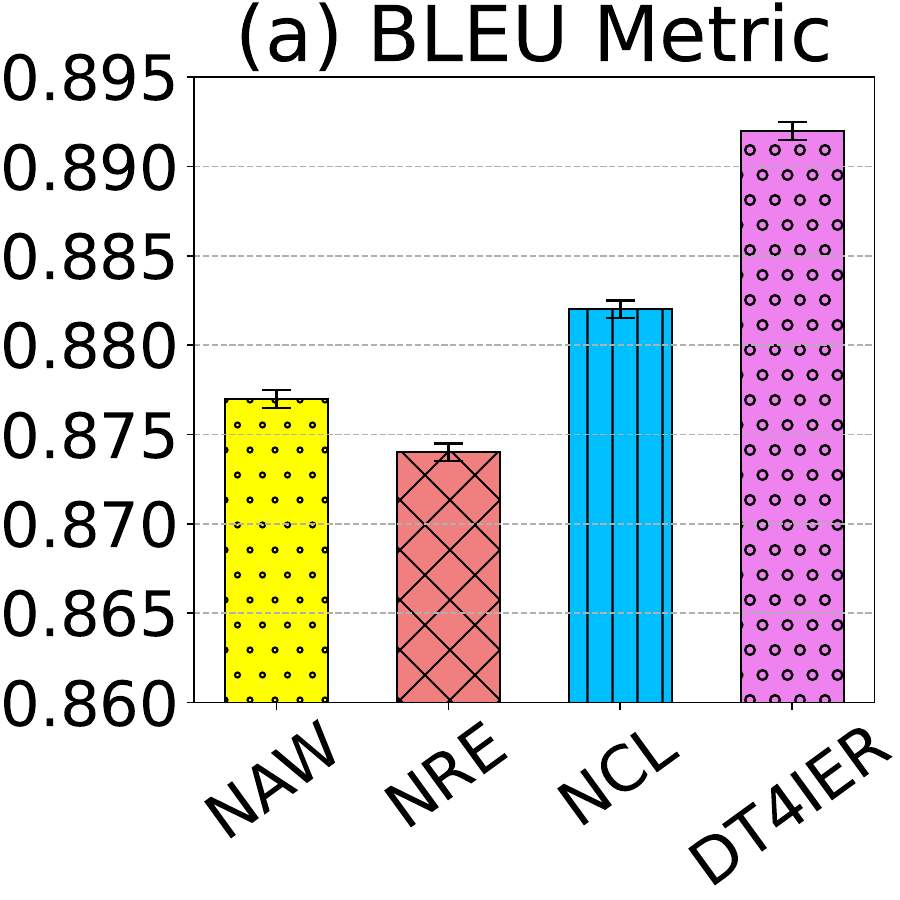}}}
	{\subfigure{\includegraphics[width=0.325\linewidth]{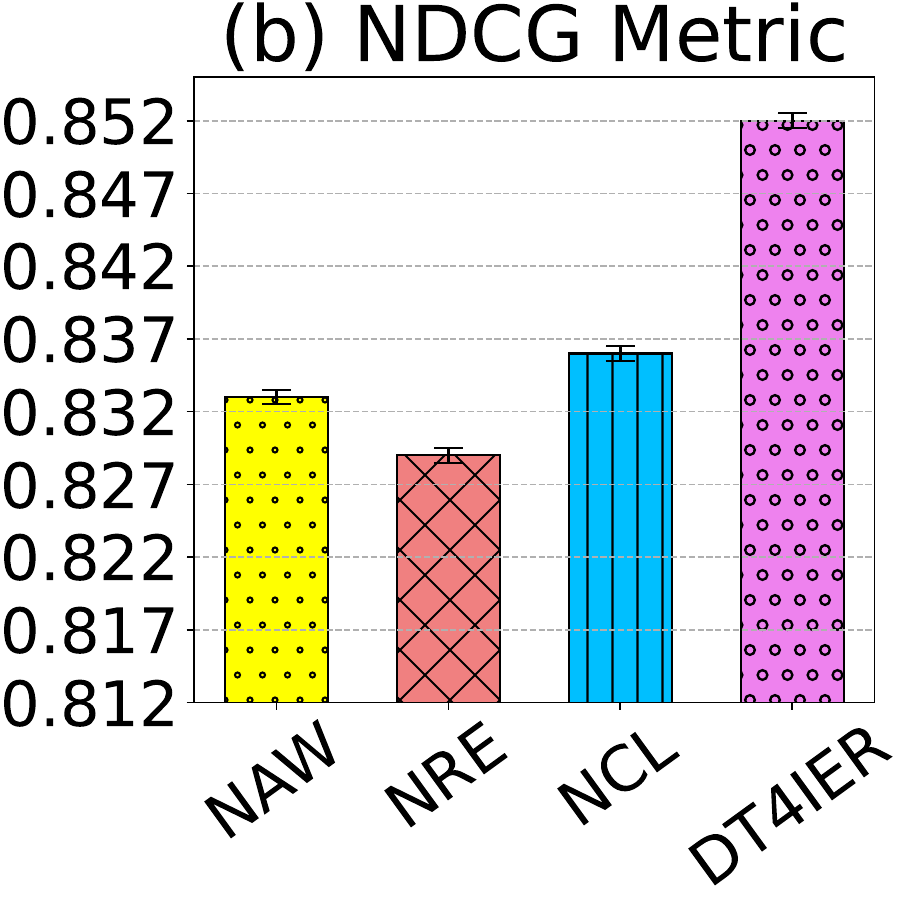}}}
	{\subfigure{\includegraphics[width=0.325\linewidth]{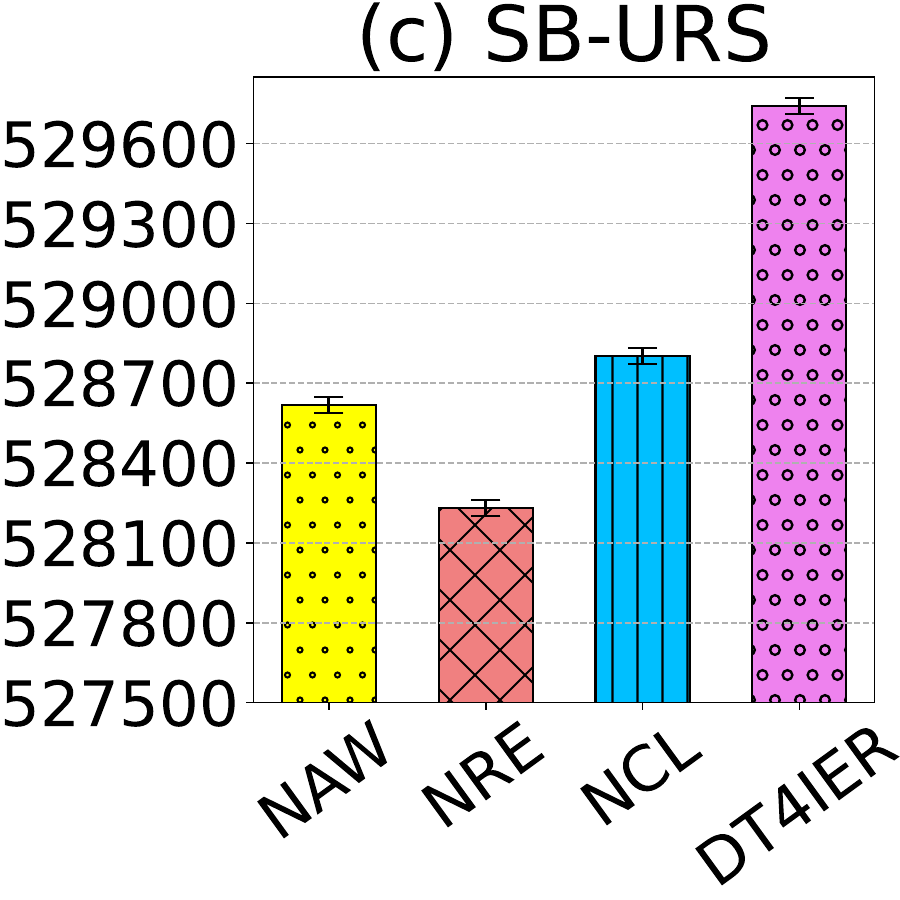}}}
    \vspace{-6mm}
    \caption{Ablation Study Results.}
    \label{Figure:2}
    \vspace{-5mm}
\end{figure}

\noindent The outcomes of our ablation study, conducted on the \name model utilizing the Kuairand-Pure dataset, are illustrated in Figure \ref{Figure:2}. From the results, we have several key insights:
\begin{itemize} [leftmargin=*]
    \item \textbf{Importance of RTG Balancing:} Our \name model consistently outperforms the NAW variant across diverse metrics, spanning recommendation accuracy to long-term retention metrics. Specifically, it achieves improvements of 1.50\% in BLEU, 1.90\% in NDCG, and 0.08\% in SB-URS. This can be largely attributed to the RTG balancing module, which adaptively weights the immediate reward and long-term retention by distinct user features. The result effectively underscores the contribution of this module.

    \item \textbf{Limitations of NRE}: The NRE configuration achieves the most modest performance across both immediate feedback and long-term retention metrics. Compared with the NRE variant, the \name model achieves improvements of 1.80\% in BLEU, 2.30\% in NDCG, and 0.19\% in SB-URS. Its primary shortcoming arises from its encoder module, which struggles to efficiently map 2-D rewards or discern the intrinsic connections between tasks. This underscores the efficiency of our proposed multi-reward embedding module in driving better performance.

    \item \textbf{Impact of Contrastive Loss}: The absence of contrastive loss in the NCL variant notably diminishes its performance. This is mainly because, without this component, the action embeddings for distinct rewards aren't adequately separated. The \name model achieves improvements of 1.02\% in BLEU, 1.60\% in NDCG, and 0.03\% in SB-URS against NCL. Notably, while the SB-URS metrics between \name and the NCL variant are comparable, our model manages to boost recommendation accuracy without compromising on long-term retention capabilities.
\end{itemize}

\subsection{RTG Prompting Analysis}
\label{RPA}
\begin{figure}[h]
    \centering    
    \vspace{-1mm}{\subfigure{\includegraphics[width=0.45\linewidth]{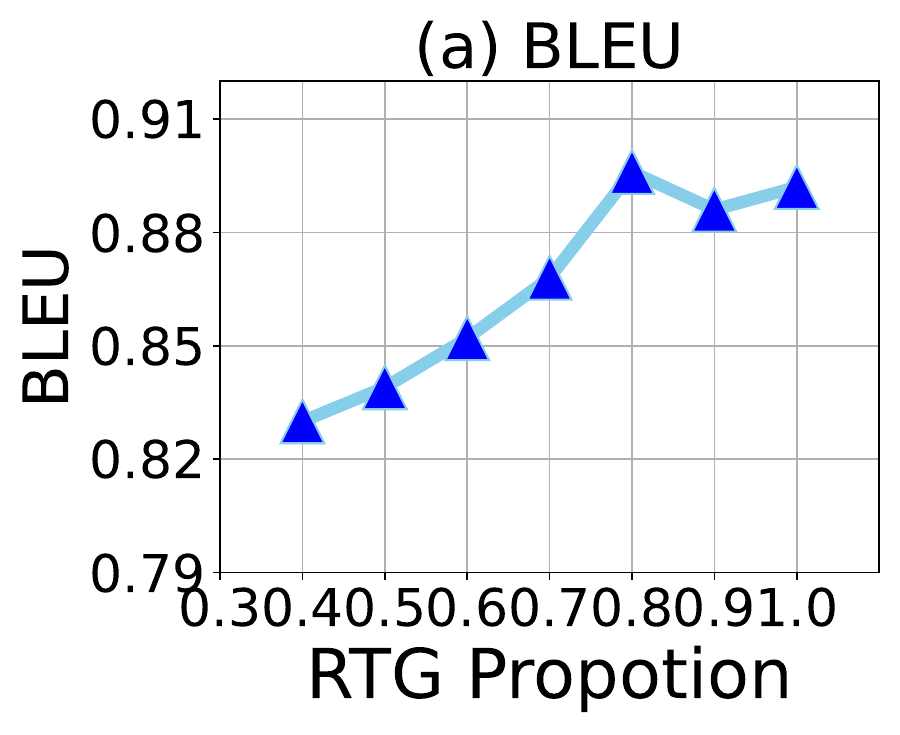}}}
	{\subfigure{\includegraphics[width=0.45\linewidth]{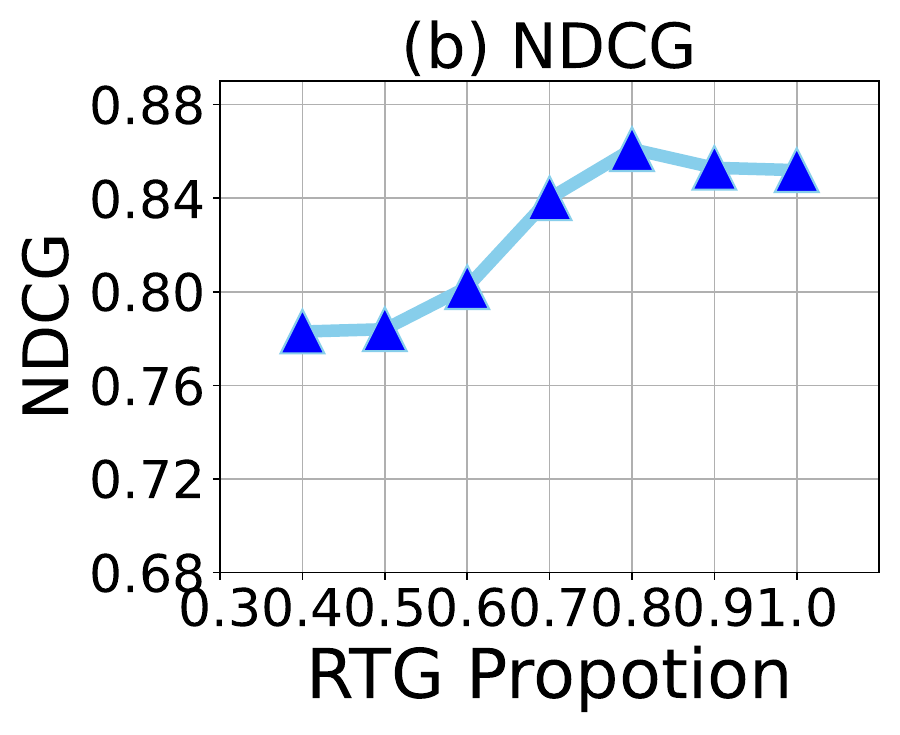}}}
    \vspace{-6mm}
    \caption{RTG Prompting Analysis.}
    \label{Figure:RTG}
    \vspace{-3mm}
\end{figure}

The inference mechanism of the Decision Transformer (DT) operates on the principle of supervised action prediction, conditioned on the highest possible Return-to-Go (RTG) values. In our specific context, this RTG value is represented as [1,1], implying an anticipated 100\% click rate along with the expectation that the user will return in the subsequent session. However, in practical applications, this ideal scenario isn't always achieved which underscores the potential benefit of utilizing RTG prompting with a reduced proportion to potentially boost model performance. Drawing from this observation, we evaluate the model's performance over RTG values ranging from 0.4 to the upper limit of 1.0, with 1.0 representing the utilization of the maximum RTG. The results are presented in Figure \ref{Figure:RTG}, offering insights into the interplay between RTG proportions and recommendation efficacy.

From the figure presented, a distinct pattern emerges in the performance metrics. Notably, both the BLEU and NDCG scores exhibit a consistent ascent with increasing RTG proportions, predominantly in the 0.4 to 0.8 range. The best performance for these metrics is achieved at an RTG proportion of 0.8, beyond which no incremental benefit is observed. This pattern suggests that while higher RTG promptings enhance recommendation accuracy, there exists a saturation point beyond which further increments do not translate to performance gains. The result exactly validates our hypothesis regarding RTG prompting.

\vspace{-2mm}

\subsection{Case Study}
In this subsection, we highlight \name's effectiveness in improving recommendation performance through a case study on the Kuairand-Pure dataset. The selected user for this study has a history of four interactions, with the ground truth action sequence comprising details of five different videos. Figure \ref{Figure:2} shows that when provided with the user's state and the goal of maximizing retention, DT4Rec recommends four videos. This recommendation achieves a BLEU score of 0.625 and an NDCG of 0.710. In contrast, \name outperforms by recommending five actions, all of which align with the ground truth action sequence. This leads to a higher prediction accuracy, with a BLEU score of 0.887 and an NDCG of 0.947. These results underscore our model's enhanced efficiency, attributed to its consideration of short-term clicks and long-term retention.

\begin{figure}[h]
    \flushleft
    \includegraphics[width=1\linewidth]{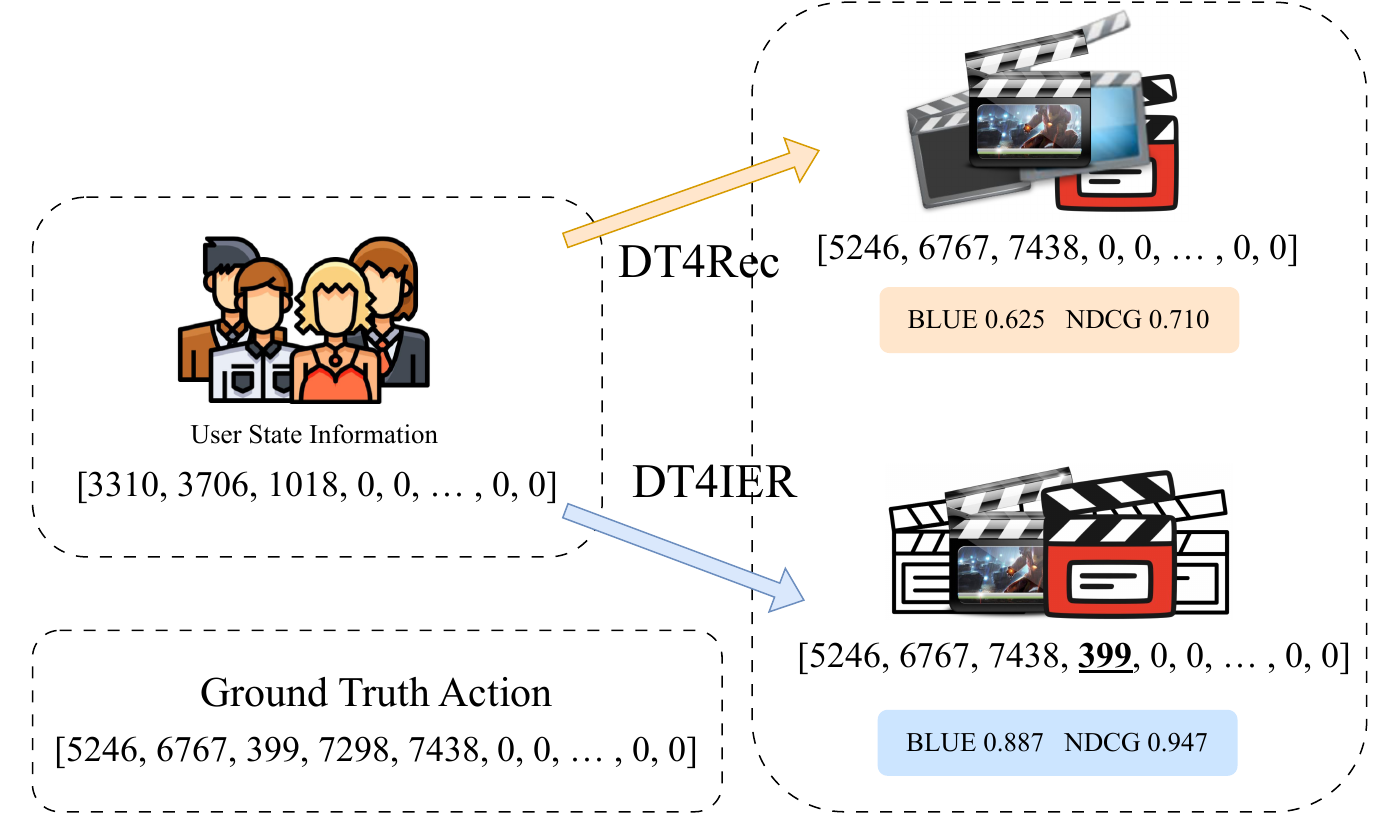}
    \vspace{-4mm}
    \caption{Case Study on Kuairand-Pure}
    \label{Figure:3}
    \vspace{-5mm}
\end{figure}

\section{Related Work}\label{sec: related_work}
In this section, we briefly discuss existing research related to sequential recommender systems, RL-based recommender systems, and MTL-based recommender systems.

\subsection{\textbf{Sequential Recommender Systems}}

Sequential Recommendation refers to a recommendation system paradigm that models patterns of user behavior and items over time to suggest relevant products or content \cite{li2022mlp4rec, zhang2024ssdrec, gao2024smlp4rec, liu2023diffusion,liu2023linrec}. Among the myriad of approaches available, Markov Chains (MCs) \cite{rendle2010factorizing} and Recurrent Neural Networks (RNNs) stand out for their prowess in sequence modeling \cite{donkers2017sequential}. The adaptability and capability of RNNs in integrating diverse information make them especially effective for crafting sequential recommendations. RU4Rec \cite{tan2016improved} was a pioneer in leveraging RNNs, tapping into their potential to process sequential data. Subsequently, GRU4Rec \cite{hidasi2016parallel} unveiled a parallel RNN structure to process item features, thus boosting recommendation quality. Further, the generalization power of the transformer architecture has given rise to its prominence in Sequential Recommendation Systems, birthing models such as BERT4Rec \cite{sun2019bert4rec} and SASRec \cite{kang2018self}. Building upon these foundations, the decision transformer was conceptualized to tackle user retention challenges, aiming to directly predict actions using a reward-driven autoregressive framework \cite{zhao2023user}. However, a gap remains in current research concerning the optimization of multi-reward settings.

\subsection{\textbf{Reinforcement Learning Based Recommender Systems}}

The validity of applying RL-based solutions~\cite{sutton2018reinforcement,afsar2021reinforcement,wang2022surrogate,zhang2022multi, zhao2018deep, zhao2021dear} for sequential recommendation comes from the assumption of Markov Decision Process~\cite{shani2005mdp}.
And the key advantage of RL solutions is the ability to improve the expected cumulative reward of future interactions with users, rather than optimizing the one-step recommendation. 
Specifically, for scenarios with small recommendation spaces, one can use tabular-based~\cite{mahmood2007learning,moling2012optimal} or value-based methods ~\cite{taghipour2007usage,zheng2018drn,zhao2018recommendations,ie2019slateq} to directly evaluate the long-term value of the recommendation;
For scenarios where action spaces are large, policy gradient methods~\cite{sun2018conversational,chen2019top,chen2019large, zhao2020jointly} and actor-critic methods~\cite{sutton1999policy,peters2008natural,bhatnagar2007incremental,degris2012model,dulac2015deep,liu2018deep,liu2020state,xin2020self, zhao2020whole,cai2023two, liu2024autoassign+} are adopted to guide the policy towards better recommendation quality. Knowing that the web service may want to optimize multiple metrics, several works have discussed the challenge of multi-objective optimization~\cite{chen2021generative,cai2023two} where the user behaviors might have different distributional patterns.
Among all user feedback signals, user retention has been considered one of the most challenging to optimize, while recent work has shown a possible RL-based approach~\cite{cai2023reinforcing} for this uphill struggle.
This work aims to simultaneously optimize immediate feedback and user retention.
Similar to our work, to overcome the gap between experiments on real user environments and offline evaluations, user simulators are widely used to bypass this paradox~\cite{ie2019recsim,zhao2019deep,zhao2023kuaisim}. Our approach can be thought of as a likelihood-based method while employing a sequence modeling objective rather than relying on variational techniques.

\subsection{\textbf{Multi-task Learning in Recommender Systems}}

Multi-task learning (MTL) is a machine learning technique that addresses multiple tasks simultaneously ~\cite{zhang2021survey, wang2023multi}. It captures a shared representation of the input through a shared bottom layer and then processes each task using distinct networks with task-specific weights. The overall performance is further enhanced by the knowledge transfer between tasks. This approach has become particularly popular in recommender systems, thanks to its prowess in effectively sharing data across tasks and in recognizing a range of user behaviors~\cite{ma2018esmm,lu2018coevolutionary,hadash2018rank,pan2019mandarin,pei2019value}. A significant portion of recent research targets the enhancement of these architectures to promote more effective knowledge sharing. Notably, some innovations focus on introducing constraints to task-specific parameters~\cite{duong2015low,misra2016cross,yang2016deep}, while others aim to clearly demarcate shared from task-specific parameters~\cite{ma2018mmoe, tang2020ple}. The principal goal of these strategies is to amplify knowledge transfer through improved feature representation. Additionally, some researchers are exploring the potential of reinforcement learning to enhance the MTL model by adjusting loss function weights \cite{liu2023multi}. While many approaches prioritize item-wise modeling, our research delves into the challenges of balancing both short-term and long-term rewards in sequential recommendation settings. Furthermore, we have refined the methodologies of both the PLE and MMoE models for sequential recommendations.

\section{Conclusion}
In this work, we introduced \name, an advanced decision transformer tailored for sequential recommendations within a multi-task framework. Our primary objective was twofold: augmenting recommendation accuracy and ensuring a harmonious balance between immediate user responses and long-term retention. To achieve this, \name leverages an innovative multi-reward mechanism that adeptly integrates immediate user responses with long-term retention signals, tailored by user-specific attributes. Furthermore, the reward embedding module is enriched by a high-dimensional encoder that deftly navigates the intricate relationships between different tasks. Empirical results on three business datasets consistently positioned \name ahead of prevalent baselines, highlighting its transformative potential in recommender systems.

\section*{ACKNOWLEDGEMENTS}
This research was partially supported by Kuaishou, Research Impact Fund (No.R1015-23), APRC - CityU New Research Initiatives (No.9610565, Start-up Grant for New Faculty of CityU), CityU - HKIDS Early Career Research Grant (No.9360163), Hong Kong ITC Innovation and Technology Fund Midstream Research Programme for Universities Project (No.ITS/034/22MS), Hong Kong Environmental and Conservation Fund (No. 88/2022), and SIRG - CityU Strategic Interdisciplinary Research Grant (No.7020046, No.7020074).

\begin{appendix}

\section{HYPER-PARAMETER SELECTION} \label{AppendixA}
In our study, we undertook a hyperparameter tuning process for the \name model to ensure the robustness and credibility of our experimental results. These results represent the aggregated performance across two datasets. A comprehensive breakdown of the hyperparameter choices is provided in Table \ref{tab:hyperparams} for further reference.

\begin{table} [h]
    \centering
    \caption{Hyper-parameter Selection for \name}
    \vspace{-3mm}
    \label{tab:hyperparams}
    \begin{tabular}{lcc}
        \toprule
        Hyper-parameter & Tuning range & Our choice \\
        \midrule
        Trajectory length & [10,20,30,40] & 20 \\
        Number of heads & [2,4,8,16] & 8 \\
        Embedding size & [64,128,256,512] & 128 \\
        LR (Kuairand) & [0.002,0.005,0.02,0.05] & 0.005 \\
        LR (ML-25M) & [0.002,0.005,0.02,0.05] & 0.02 \\
        LR (RetailRocket) & [0.002,0.005,0.02,0.05] & 0.02 \\
        balance term $\gamma$ & [0.2,0.3,0.4,0.5,0.6,0.7,0.8] & 0.5 \\
        Loss weight $\alpha$ & [0.05,0.1,0.2,0.3] & 0.1 \\
        \bottomrule
    \end{tabular}
    \vspace{-1mm}
\end{table}

\section{Other Model details}

\subsection{State-action Encoder} \label{S-A E}
Given the state and action sequence of length $H=N$, we apply GRU to process the input sequence and specify by state $\textbf{s}_t$:

\begin{equation} \label{sae}
\begin{aligned}
    z_n & =\sigma'\left(W_z \textbf{s}_{t,n}+U_z h_{n-1}+b_z\right) \\
    o_n & =\sigma'\left(W_o \textbf{s}_{t,n}+U_o h_{n-1}+b_o\right) \\
    h_n & = f(h_{n-1},z_n) \\
    \widehat{\textbf{E}}^{s}_t & = h_N
\end{aligned}
\end{equation}
where $z_n, o_n$ are update gate vector and reset gate vector for $n$-th item, $h_n$ is the output vector, $\sigma'$ is the logistic function, $W, U, b$ are parameter matrices and vector, $\widehat{\textbf{E}}^{s}_t$ is the state embedding at $t$ timestamp which is also the last output vector for $N$.

\subsection{Transformer Block}
\label{TB}
We employ a unidirectional transformer layer equipped with a multi-head self-attention mechanism as our primary model architecture. To combat overfitting, skip-connections are integrated, and feed-forward neural layers are utilized for feature transformation:
\begin{equation}
    \widehat{\textbf{E}}^{A}=\text { FFN }\left[\text { MHA }\left(\boldsymbol{\tau}^{\prime}\right)\right]
\end{equation}
where $\widehat{\textbf{E}}^{A}$ is the predicted action embedding, $\boldsymbol{\tau}^{\prime}$ is the trajectory information containing state-action and RTG embedding. $\text { MHA }$ is a multi-head self-attentive layer and $\text{FNN}$ is the feed-forward neural with Gaussian Error Linear Units (GELU) activation function.

\subsection{Action Decoder}
\label{AD}
Given predicted action embedding $\widehat{\textbf{E}}^{A}$ with $t$-th rows as $A_t$ and user interaction histories $i_n$, the action decoder aims to decode sequences of items of interest to users with the GRU module:
\begin{equation}
\begin{aligned}
    \widehat{i}_n & =i_n \oplus A_t \\
    \widehat{h}_{n+1} & =f(\widehat{i}_n, \widehat{h}_{n})
\end{aligned}
\end{equation}
where $\widehat{h}_{n}$ is the $n$-th output vector. To forecast the first item without any prior information, we employ 'start' as a checkpoint and initialize $\widehat{i}_0$ arbitrarily. The decoding process can be expressed as:

\begin{equation}
\begin{aligned}
    \widehat{\textbf{a}}_t & = \text{decode}( \text{start}, \widehat{i}_1, \ldots, \widehat{i}_{N-1}) \\
    & = [ \widehat{h}_{1}, \ldots, \widehat{h}_{N} ]
\end{aligned}
\end{equation}

\noindent where $\widehat{\textbf{a}}_t$ represents the predicted action for $t$-th instance.

\end{appendix} \twocolumn \balance

\bibliographystyle{ACM-Reference-Format}
\balance
\bibliography{ref}

\end{document}